\documentclass[aps,prc,twocolumn,showpacs,superscriptaddress]{revtex4}
\usepackage{graphicx}
\usepackage{hyperref}
\usepackage{amssymb}

\newcommand{\Aproj}{\ensuremath{A_\mathrm{proj}}}
\newcommand{\dndy}{\ensuremath{dN/dy}}
\newcommand{\Ebeam}{\ensuremath{E_\mathrm{beam}}}
\newcommand{\EZCAL}{\ensuremath{E_\mathrm{ZCAL}}}
\newcommand{\fZCAL}{\ensuremath{f_\mathrm{ZCAL}}}
\newcommand{\NINT}{\ensuremath{N_\mathrm{INT}}}
\newcommand{\npp}{\ensuremath{N_\mathrm{pp}}}
\newcommand{\ncoll}{\ensuremath{N_\mathrm{coll}}}
\newcommand{\sigint}{\ensuremath{\sigma_\mathrm{tot}}}
\newcommand{\sNN}{\ensuremath{\sqrt{s_{NN}}}}
\newcommand{\nppZCAL}{\ensuremath{N_\mathrm{pp}^\mathrm{ZCAL}}}
\newcommand{\TB}{\ensuremath{T_\mathrm{B}}}

\newcommand{\AGeV}{\ensuremath{A\,\mathrm{GeV}}}
\newcommand{\AGeVc}{\ensuremath{A\,\mathrm{GeV}/c}}
\newcommand{\barns}{\ensuremath{\mathrm{b}}}
\newcommand{\GeV}{\ensuremath{\mathrm{GeV}}}
\newcommand{\GeVc}{\ensuremath{\mathrm{GeV}/c}}
\newcommand{\GeVcc}{\ensuremath{\mathrm{GeV}/c^2}}
\newcommand{\mb}{\ensuremath{\mathrm{mb}}}
\newcommand{\MeV}{\ensuremath{\mathrm{MeV}}}
\newcommand{\MeVcc}{\ensuremath{\mathrm{MeV}/c^2}}

\newcommand{\PKp}{\ensuremath{K^+}}
\newcommand{\PKm}{\ensuremath{K^-}}

\newcommand{\PKS}{\ensuremath{K_S}}
\newcommand{\Pphi}{\ensuremath{\phi}}

\newcommand{\mean}[1]{\ensuremath{{\left<#1\right>}}}
\newcommand{\ppcoll}{\ensuremath{p+p}}
\newcommand{\pncoll}{\ensuremath{p+n}}
\newcommand{\nncoll}{\ensuremath{n+n}}
\newcommand{\pAcoll}[1]{\ensuremath{p+\mathrm{#1}}}
\newcommand{\coll}[2]{\ensuremath{\mathrm{#1}+\mathrm{#2}}}

\newcommand{\Eq}[1]{Eq.~(\ref{#1})}
\newcommand{\about}[1]{\mbox{\ensuremath{\sim}\ensuremath{#1}}}

\begin{document}

\title{Production of \Pphi\ Mesons in \coll{Au}{Au} Collisions at 11.7\AGeVc}



\newcommand {\ANL} {\mbox{Argonne National Laboratory, Argonne, IL
60439, USA}}

\newcommand {\BNL} {\mbox{Brookhaven National Laboratory, Upton, NY
11973, USA}}

\newcommand {\UCR} {\mbox{University of California, Riverside, CA
92521, USA}}

\newcommand {\CU} {\mbox{Columbia University, Nevis Laboratories,
Irvington, NY 10533, USA}}

\newcommand {\UI} {\mbox{University of Illinois at Chicago, Chicago,
IL 60607, USA}}

\newcommand {\UMD} {\mbox{University of Maryland, College Park, MD
20742, USA}}

\newcommand {\MIT} {\mbox{Massachusetts Institute of Technology,
Cambridge, MA 02139, USA}}

\newcommand {\UR} {\mbox{University of Rochester, Rochester, NY 14627,
USA}}

\newcommand {\YONSEI}{\mbox{Yonsei University, Seoul 120-749, South
Korea}}

\newcommand {\Wen} {Present address: Institute of Physics, Academia
Sinica, Taipei 11529, Taiwan.}

\newcommand {\Jamie} {Present address: Brookhaven National Laboratory,
Upton, NY 11973, USA.}

\newcommand {\Albrecht} {Present address: Forschungzentrum
J\"{u}lich, J\"{u}lich, D-52425, Germany.}

\newcommand {\Walter} {Present address: Gesellschaft f\"{u}r
Schwerionenforschung, D-64291 Darmstadt, Germany.}

\newcommand {\Dave} {Present address: University of Illinois at Chicago,
Chicago, IL 60607.}

\newcommand {\Moulson} {Present address: Laboratori Nazionali di
Frascati dell'INFN, 00044 Frascati RM, Italy.}

\newcommand {\Vandana} {Present address: Tata Institute of
Fundamental Research, Colaba, Mumbai 400005, India.}

\newcommand {\Craig} {Present address: Iowa State University, Ames, IA 50011.}

\newcommand {\Dan} {Present address: National Institute of Health,
Gaithersburg, MD 20892.}

\newcommand {\Wang} {Present address: CW Associates, 7676 Woodbine,
Markham L3R 2N2, Ontario.}

\newcommand {\Zou} {Present address: H.-M. Zou, Rabobank Nederland, 245 Park Ave, New York, NY 10167.}

\author{B.B.~Back}\affiliation{\ANL}
\author{R.R.~Betts}\affiliation{\ANL}\affiliation{\UI}
\author{J.~Chang}\affiliation{\UCR}
\author{W.C.~Chang}\altaffiliation{\Wen}\affiliation{\UCR}
\author{C.Y.~Chi}\affiliation{\CU} \author{Y.Y.~Chu}\affiliation{\BNL}
\author{J.B.~Cumming}\affiliation{\BNL}
\author{J.C.~Dunlop}\altaffiliation{\Jamie}\affiliation{\MIT}
\author{W.~Eldredge}\affiliation{\UCR}
\author{S.Y.~Fung}\affiliation{\UCR}
\author{R.~Ganz}\affiliation{\UI}
\author{E.~Garcia}\affiliation{\UMD}
\author{A.~Gillitzer}\altaffiliation{\Albrecht}\affiliation{\ANL}
\author{G.~Heintzelman}\altaffiliation{\Jamie}\affiliation{\MIT}
\author{W.F.~Henning}\altaffiliation{\Walter}\affiliation{\ANL}
\author{D.J.~Hofman}\altaffiliation{\Dave}\affiliation{\ANL}
\author{B.~Holzman}\altaffiliation{\Jamie}\affiliation{\UI}
\author{J.H.~Kang}\affiliation{\YONSEI}
\author{E.J.~Kim}\affiliation{\YONSEI}
\author{S.Y.~Kim}\affiliation{\YONSEI}
\author{Y.~Kwon}\affiliation{\YONSEI}
\author{D.~McLeod}\affiliation{\UI}
\author{A.C.~Mignerey}\affiliation{\UMD}
\author{M.~Moulson}\altaffiliation{\Moulson}\affiliation{\CU}
\author{V.~Nanal}\altaffiliation{\Vandana}\affiliation{\ANL}
\author{C.A.~Ogilvie}\altaffiliation{\Craig}\affiliation{\MIT}
\author{R.~Pak}\altaffiliation{\Jamie}\affiliation{\UR}
\author{A.~Ruangma}\affiliation{\UMD}
\author{D.E.~Russ}\altaffiliation{\Dan}\affiliation{\UMD}
\author{R.K.~Seto}\affiliation{\UCR}
\author{P.J.~Stanskas}\affiliation{\UMD}
\author{G.S.F.~Stephans}\affiliation{\MIT}
\author{H.Q.~Wang}\altaffiliation{\Wang}\affiliation{\UCR}
\author{F.L.H.~Wolfs}\affiliation{\UR}
\author{A.H.~Wuosmaa}\affiliation{\ANL}
\author{H.~Xiang}\affiliation{\UCR} 
\author{G.H.~Xu}\affiliation{\UCR}
\author{H.B.~Yao}\affiliation{\MIT} 
\author{C.M.~Zou}\altaffiliation{\Zou}\affiliation{\UCR}

\collaboration{E917 Collaboration}


\begin{abstract}
We report on a measurement of \Pphi-meson production in \coll{Au}{Au}
collisions at a beam momentum of 11.7\AGeVc\ by Experiment 917 at the
AGS. The measurement covers the midrapidity region $1.2 < y < 1.6$.
Transverse-mass spectra and rapidity distributions are presented as
functions of centrality characterized by the number of participant
projectile nucleons. The yield of \Pphi's per participant projectile
nucleon increases strongly in central collisions in a manner similar
to that observed for kaons.
\end{abstract}


\maketitle

\section{Introduction}
\label{sec:intro}

The production of \Pphi\ mesons in relativistic heavy-ion collisions
has been an important subject of study at the AGS, the SPS, and RHIC.
The production of the \Pphi\ meson, the
lightest bound state of strange quarks ($s\bar{s}$), is suppressed in
ordinary hadronic interactions because of the Okubo-Zweig-Iizuka (OZI)
rule~\cite{OZI}. It has been proposed that in a quark-gluon plasma
(QGP) scenario, strange quarks could be rapidly and abundantly
produced via gluon interactions~\cite{STRANGE}. Thus, \Pphi\ mesons
could be created in a non-conventional way via strange quark
coalescence, bypassing the OZI rule. A strong enhancement in
\Pphi-meson production would serve as one of the
strangeness-enhancement signatures for QGP formation~\cite{SHOR}.

The production of \Pphi\ mesons has been measured in \coll{Si}{Au}
collisions at 14.6\AGeVc\ by E859 at the AGS within a rapidity
interval of $1.2 < y < 2.0$~\cite{E859PRL}. 
The ratio of the total \Pphi\ yield to the \PKm\ yield has been found to 
be 10\% for the uppermost 7\% of the charged-particle multiplicity
distribution.
NA49 at the SPS has measured \Pphi-meson production in
\ppcoll, \pAcoll{Pb}, and \coll{Pb}{Pb} collisions at a beam energy
($E_\mathrm{beam}$) of 158\AGeV\ within a rapidity range of $3.0 < y <
3.8$. An enhancement in the ratio of \Pphi\ to pion total yields by a
factor of $3.0 \pm 0.7$ has been observed in central \coll{Pb}{Pb}
collisions relative to in \ppcoll\ interactions~\cite{NA49PHI}.
Also at the SPS, NA50 has measured the ratio $\Pphi/(\rho+\omega)$ as 
a function of centrality in \coll{Pb}{Pb} collisions by
fitting the invariant-mass spectrum of muon pairs with 
$0 < y - y_{NN} < 1$ and $1.5 < m_t \lesssim 3$ \GeVcc~\cite{NA50PHI}.
Both this ratio and the absolute yield of 
\Pphi's per participant nucleon have been found to increase significantly
with the number of participant nucleons.
The STAR experiment has
reported on a measurement of the \Pphi\ yield at midrapidity in
\coll{Au}{Au} collisions at center-of-mass energy $\sNN =
130$~\GeV. An increase in the $\Pphi/h^-$ ratio with $\sNN$ has been
observed~\cite{STARPHI}.

Enhancements in the yields of particles with open strangeness have
also been observed.  At the AGS,
$\PKp/\pi^+$ ratios have been seen to increase to \about{20}\% in central
\coll{Si}{Au} and \coll{Au}{Au} collisions, from 4--8\% in \ppcoll\
collisions~\cite{E802KPI,E866KAON,E917KPI}, and the
$\bar{\Lambda}/\bar{p}$ ratio has been found to increase strongly with
centrality in \coll{Au}{Au} collisions~\cite{E864LBAR,E917LBAR}.  At
the SPS, the WA97 collaboration has observed enhanced production of
$K$'s, $\Lambda$'s, $\Xi$'s, and $\Omega$'s in heavy-ion collisions
relative to in \ppcoll\ or \pAcoll{A} collisions~\cite{WA97S}.
Measurements from the PHENIX experiment at center-of-mass energy $\sNN
= 130$~\GeV\ have shown the kaon yield to be more strongly dependent on
centrality than the pion yield at midrapidity~\cite{PHENIXKPI}.

The reason for the strangeness enhancement in heavy-ion collisions is
not completely understood. The rescattering of hadrons and the
conversion of the excitation energy of secondary resonances into
strange particles might give rise to strangeness enhancement in a
purely hadronic picture~\cite{RQMD,ARC,ART1}. Strangeness enhancement
can also be interpreted as a reduction in canonical strangeness
suppression from $p+p$ to $A+A$ reactions in the context of thermal
models~\cite{Thermal}. Strangeness enhancement as a function of the
number of ``grey'' protons (a centrality index) has also been seen in
\pAcoll{A} collisions, in which a QGP phase is unlikely to
contribute~\cite{E910PRL}.

Under the conventional hadronic interactions, there are mainly three
scenarios proposed for the production of \Pphi\ mesons in nuclear
collisions:

\begin{itemize}

\item Parton fusion of strange sea quarks~\cite{PARTONFUSION} or
  knock-out of $s \bar{s}$ pairs~\cite{KNOCKOUT} from the primary
  collisions of projectile and target nucleons.

\item Secondary baryon-baryon interactions $BB \rightarrow \phi
  NN$, meson-baryon interactions $(\pi, \rho)B \rightarrow \phi
  B$~\cite{ART2} and meson-meson interactions $\pi \rho \rightarrow
  \phi$~\cite{RQMDPHI}, where $B = N, \Delta, N^*$. 

\item Secondary kaon-hyperon interactions $K Y \rightarrow
  \phi N$ and kaon-antikaon scattering $K \bar{K} \rightarrow \phi
  \rho$ in the event of the restoration of chiral symmetry in the hot and
  dense nuclear fireball~\cite{ART3}.

\end{itemize}
With inclusion of secondary meson-baryon and meson-meson interactions,
RQMD (Relativistic Quantum Molecular Dynamics)~\cite{RQMDPHI} is able
to qualitatively describe the increase of kaon yields with centrality
measured by E866~\cite{E859KAA}, though quantitative differences do
exist~\cite{FQWANG}.

The different scenarios described above imply different relations between 
the scaling of \Pphi\ production and that of other hadrons, such as 
kaons and pions, with collision centrality and energy. 
Hence, a systematic measurement of
\Pphi\ production in different collisional systems may help to
quantify the increase in the overall strangeness production, and, in
combination with other measurements of strange and non-strange hadron
production, to differentiate between possible mechanisms contributing
to the strangeness enhancement~\cite{Dunlop}.

In this paper we report on a measurement of the \Pphi\ yield around
midrapidity in \coll{Au}{Au} collisions at the Alternating Gradient
Synchrotron (AGS) at Brookhaven National Laboratory (BNL), and compare
it to the yields of pions and kaons.

\section{Experimental Details}
\label{sec:expt}

Experiment 917 took data on \coll{Au}{Au} reactions at projectile
momenta of 6.8, 8.9 and 11.7\AGeVc\ (corresponding to center-of-mass
energies $\sNN$ = 3.83, 4.31, and 4.87~\GeV, respectively) in the fall of
1996. The identified particles include particles containing strange
quarks, such as \PKp, \PKm, \Pphi, $\Lambda$, and $\bar{\Lambda}$, and
non-strange particles such as $\pi$, $p$, and
$\bar{p}$~\cite{E917KPI,E917LBAR,E917KPKM,E917PRO,E917PROC}.  The
experimental apparatus consisted of a movable magnetic spectrometer
for tracking and particle identification, and beamline detector arrays
for global event characterization. When the data reported in this work
were collected, the beam momentum was 11.7\AGeVc, an Au target
with areal density 1961 mg/cm$^2$ (corresponding to approximately 
4\% of an interaction length for an Au projectile) was used, and the
spectrometer angle was set to either $14^\circ$ or $19^\circ$ from the
beam axis. The momentum resolution of the spectrometer, $\delta p/p$,
was about 1\% for particles with momentum greater than $1$~\GeVc\ and
increased at lower momentum, up to 2\% at $p=0.6$~\GeVc, due to the
effect of multiple scattering. The kaon momenta in reconstructed
\Pphi\ events lay mostly in the range of $1.0 < p < 2.0$~\GeVc. Event
centrality was characterized by the energy of the beam spectators
after the interaction as measured in the zero-degree calorimeter
(ZCAL), positioned downstream of the target on the beam axis. More
details on the detector systems are given in
Refs.~\cite{E917PROC,E802,E917PHD}.

The data presented here were collected using a two-level online
trigger: a minimum-bias spectrometer-activity trigger (LVL1) followed by a
particle-identification trigger (LVL2) which required two charged
kaons of either sign or one $\bar{p}$ in the spectrometer
acceptance. The hardware LVL2 trigger looped over the combinations of
drift chamber and TOF hits, and formed combinations consistent with
tracks corresponding to particles of given momentum, charge, and
particle type using a look-up table~\cite{LVL2}. The LVL2 trigger
increased the live time of the data-acquisition system by essentially the
ratio of the rate of vetoable LVL1 triggers to the rate of vetoable
LVL1 triggers that were not vetoed. This ratio was known as the LVL2
rejection factor, and could be estimated online. In \coll{Au}{Au} collisions
with a magnetic field of 4~KG and the spectrometer at $14^\circ$ and
$19^\circ$, typical rejection factors for a $2K/\bar{p}$ trigger
were $4.8$ and $8.5$, respectively. By examining data that was taken
with the LVL1 trigger and recording the decision of the LVL2 trigger,
the inefficiency of the latter was monitored and found to be less than 
1\% for events fully inside the acceptance. Most of the LVL2-triggered
data was background since the trigger was optimized to reject only events that
were clearly not of the correct type in order to keep its efficiency near
100\%.

A time-of-flight (TOF) system with a typical resolution of 130~ps
served to identify pions and kaons up to a momentum of 1.75~\GeVc\
with 3 standard deviations of TOF resolution. Above this momentum, the
$3\sigma$ contours in the ($p$, TOF) plane began to overlap. Within
this region, particles were identified as kaons only if their TOF was
inside the kaon region and outside of the pion region, since 10--30\%
of the tracks in the overlap region were from pions. 
A specific correction for the kaon-identification inefficiency from the 
exclusion of the pion bands in this region of high momenta was implemented
in the analysis (see Sec.~\ref{sec:yields}). Additionally,
kaons with momenta less than 0.5~\GeVc\ were rejected, in order to avoid the
need for large acceptance corrections.

\begin{table}
\caption
{Centrality bins used in the analysis. The cuts on zero-degree energy,
  \EZCAL, that define each bin are listed (in \GeV), together with the
  corresponding fraction of the total cross section for
  \coll{Au}{Au} collisions (in percent), the mean number of projectile
  participants for the bin, \mean{\npp}, and the estimated value of
  the mean number of binary collisions, \mean{\ncoll}, by Glauber
  model and the mean number of projectile participants estimated from
  \EZCAL, \mean{\nppZCAL}. The total beam kinetic energy in the
  collisions is about $2123$~\GeV.}
\begin{ruledtabular}
\begin{tabular*}{0.48\textwidth}{c c c c c c}
bin & \EZCAL\ & \% \sigint\  & \mean{\npp} & \mean{\ncoll} & \mean{\nppZCAL} \\ \hline
1 & 0-280    & 0-5   & $170 \pm 10$ & $770 \pm 72$ & $179 \pm 1.0$ \\
2 & 280-560  & 5-12  & $138 \pm 12$ & $585 \pm 74$ & $157 \pm 1.5$ \\
3 & 560-960  & 12-23 & $101 \pm 14$ & $393 \pm 74$ & $126 \pm 2.0$ \\
4 & 960-1440 & 23-39 & $ 62 \pm 14$ & $207 \pm 60$ & $ 84 \pm 2.6$ \\
5 & $>$1440  & 39-77 & $ 20 \pm 12$ & $ 50 \pm 38$ & $ 34 \pm 3.1$ \\
\end{tabular*}
\end{ruledtabular}
\label{tab:cent}
\end{table}

The index of event centrality was obtained from the distribution of the
energy deposited in the zero-degree calorimeter, \EZCAL, obtained using
a minimum-bias interaction trigger based on the measurement
with a \v{C}erenkov counter of the
projectile charge after passage through the target.
Empty-target runs were periodically taken in order to subtract 
the background from the ZCAL energy distribution.
Assuming a monotonic relation between the energy deposition in the
calorimeter and the event centrality such that the most central events
correspond to the smallest energy deposition, we divided the data
into five centrality bins. The centrality bins are expressed as a fraction
of the total cross section for \coll{Au}{Au} 
collisions, 6.8~\barns\ as evaluated from the parameterizations of 
Ref.~\cite{Geer}. 
The cross section for \coll{Au}{Au} collisions as measured with the
minimum-bias interaction trigger is approximately 5.2~\barns, 
or 77\% of the total cross section. 

We have attempted to estimate \mean{\npp}, the mean
number of projectile participants, and \mean{\ncoll}, the mean number
of binary collisions, according to the Glauber model~\cite{Glauber},
using impact parameter cuts corresponding to the centrality range for
each bin as listed in Table~\ref{tab:cent}, and an assumed \ppcoll\ 
inelastic cross section of 30~\mb. In Experiments 802, 856 and 
866~\cite{E866KAON}, 
the number of projectile participants has been conventionally estimated
from the relation $\nppZCAL = \Aproj(1 - \EZCAL/\Ebeam)$, where $\Aproj$ 
is the atomic mass of the projectile, and \Ebeam\ is the total kinetic energy
of the nuclei in the beam before interaction, which was 2123 \GeV\ in
the present case. For the purpose of comparison, we have also calculated
\nppZCAL\ and its uncertainty with a nominal ZCAL energy resolution
($\sigma_{E}/E$) of 3.6\%~\cite{E802ZCAL}. All relevant parameters for
each centrality bin are listed in Table~\ref{tab:cent}. To facilitate
comparison with other experiments we will use \mean{\npp} as calculated
using the Glauber model for the rest of this paper.

\section{Experimental Results}

\begin{figure}
\includegraphics[width=0.48\textwidth]{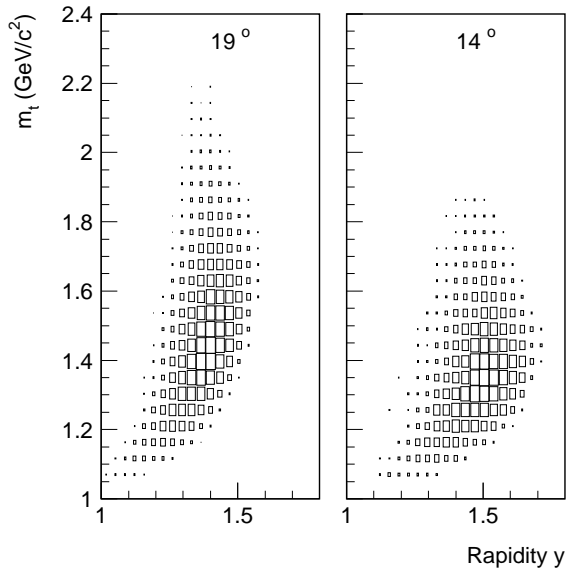}
\caption
{The experimental acceptance for \PKp\PKm\ pairs in the space of
transverse mass ($m_t$) vs.\ rapidity ($y$) for the 19$^\circ$ and 14$^\circ$
spectrometer angle settings.}
\label{fig:accept}
\end{figure}

A total of 250 million LVL2-triggered events were analyzed. When the
two spectrometer angle settings were combined, the overall acceptance
for \PKp\PKm\ pairs covered the region of rapidity $1.2 < y < 1.6$ and
transverse mass $1.0~\GeVcc < 
m_t \equiv [m_{\phi}^2+(p_t/c)^2]^{1/2}
< 2.2~\GeVcc$, as shown in Fig.~\ref{fig:accept}.

The \Pphi\ mesons were reconstructed by forming the invariant-mass
($m_{inv}$) distribution of identified \PKp\PKm\ pairs and
subtracting the combinatorial background, which was obtained by the
event-mixing method~\cite{MIXING}. Two kinds of distributions were
formed: ``same-event'' distributions, in which the \PKp's and \PKm's
were selected from the same event, and ``mixed-event'' distributions,
in which the individual particles were chosen from different events in
the same centrality class to represent the uncorrelated background.

\subsection{Invariant-mass distribution for \PKp\PKm\ pairs}
\label{sec:minv}

First, we examined the minimum-bias invariant-mass distribution of
kaon pairs by fitting the data to a relativistic Breit-Wigner
distribution (RBW) convoluted with a Gaussian representing the
experimental mass resolution~\cite{E859PRL,JDJACKSON}. The shape of
the background (BG) was obtained from the mixed-event distribution
while the normalization was left as a free parameter in the fit. The
parameterization is as follows:
\begin{eqnarray} 
\frac{dN_{\PKp\PKm}^\mathrm{same}}{dm} &=& a \int_{m_1}^{m_2}\!
\mathrm{RBW}(m')
\frac{\exp{\left[-\frac{1}{2}(\frac{m-m'}{\sigma_m})^2\right]}}{\sqrt{{2\pi
      \sigma_m^{2}}}}\; dm' \nonumber \\ & + &
      b\,\frac{dN_{\PKp\PKm}^\mathrm{mixed}}{dm},
\label{eq:minvfit}
\end{eqnarray}
where
\begin{eqnarray} 
         \mathrm{RBW}(m) & = & \frac{ m m_0 \Gamma(m)}{(m^2-m_0^2)^2+(m_0 \Gamma(m))^2}, \\
	\Gamma(m) & = & 2 \Gamma_0 \frac{(q/q_0)^3}{(q/q_0)^2+1}, \\
	q_0 & = & \sqrt{m_0^2/4 - m_K^2}, \\
	q & = & \sqrt{m^2/4 - m_K^2},
\end{eqnarray}
and where the limits of integration were $m_1 = 0.989~\GeVcc$ and $m_2
= 1.252~\GeVcc$.  There were five free parameters in all: the relative
normalizations $a$ and $b$, the peak mass and width parameters $m_0$
and $\Gamma_0$, and the experimental mass resolution $\sigma_m$.

\begin{figure}
\includegraphics[width=0.48\textwidth]{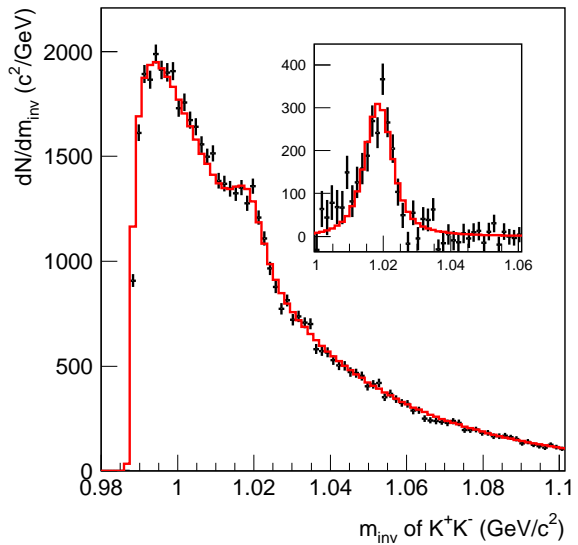}
\caption
{(Color Online) The invariant-mass ($m_{inv}$) distribution of
\PKp\PKm\ pairs for minimum-bias events, superimposed with a fit
consisting of a relativistic Breit-Wigner distribution plus the
background distribution from a mixed-event technique, as described in
the text. In the inset, the \Pphi\ signal together with fit results
from the first term in \Eq{eq:minvfit} is shown.}
\label{fig:minv}
\end{figure}

For the minimum-bias data, the same-event invariant-mass distribution
of \PKp\PKm\ pairs and the background-subtracted signal distribution
corresponding to the first term in \Eq{eq:minvfit} are shown together
with the fits in Fig.~\ref{fig:minv}. The values $m_0 =
1018.99\pm0.36$~\MeVcc, $\Gamma_0 = 6.14\pm2.59$~\MeVcc, and
$\sigma_m = 2.43\pm1.11$~\MeVcc\ were obtained with
$\chi^2/\mathrm{dof} = 145/167 = 0.87$. The fit results for the 
peak position ($m_0$) and width ($\Gamma_0$) of the \Pphi\ signal
are in agreement with the
world-average values. The large uncertainty on $\Gamma_0$ arises mostly
because of a high degree of correlation in the fit between the values
for the parameters $\Gamma_0$ and $b$ (correlation coefficient =
$0.71$). The value obtained for the experimental mass resolution ($\sigma_m$) 
is also
consistent with our estimated value of 2.0~\MeVcc, from Monte Carlo
studies and from the width of the peak from $\Lambda$ decay in the
$p\pi^-$ invariant-mass distribution~\cite{E917PHD} (the known
contribution from the multiple scattering of the kaons in the target
in the case of the present measurement was accounted for by additional
smearing). It is noted that the result of the fit lies systematically
below the data in the invariant-mass region 10--20 \MeVcc\ below the
\Pphi\ peak. We do not have a clear understanding of this effect.

\subsection{Signal counting and extraction of differential yields}
\label{sec:yields}

Monte Carlo studies confirm that the experimental invariant-mass
resolution ($\sigma_m$) remains constant (\about{2}~\MeVcc) across the
kinematic region of acceptance~\cite{E917PHD}. Given the stable mass
resolution and the limited statistics for the division of data into
bins in centrality and phase space, the transverse-mass spectra were
obtained by counting events in a defined signal window and estimating
the number of background events within this window, rather than by
integration of the result of the fit with \Eq{eq:minvfit}. The details
were as follows. The data were divided into subsets for each of the two
spectrometer settings and five centrality classes. Same-event and
mixed-event pairs for each subset were binned in rapidity and
transverse mass ($\Delta y=0.2$ and $\Delta m_t=0.2~\GeVcc$), and
invariant-mass distributions were obtained for each bin.
No corrections were applied to the invariant-mass distributions at this 
stage---signal extraction was performed on the 
spectra of raw counts.   
In each bin,
the total number of signal and background counts ($S+B$) was taken to
be the sum of the counts within a signal window defined as $1.0040 <
m_{inv} < 1.0355$ \GeVcc. In units of the natural $\phi$ width, this
interval corresponds to approximately $\pm3.7\Gamma_0$ about the expected
peak position. We then estimated the number of background
counts ($B$) inside the signal window as the number of counts in the
normalized distribution of mixed-event pairs in this same
interval. For this purpose, the mixed-event distributions were 
normalized to have the same area outside of the signal window 
as the same-event distributions.
The final number of signal counts ($S$) was obtained by subtracting the
background estimated in this way from the total number of events
in the signal window.
Using this procedure, we were able to obtain estimates of $S$ for 
a total of 88 distributions corresponding to different bins in phase space
and centrality for each of the two angle settings.

We use the $\chi^2$ statistic obtained from the Poisson log-likelihood
treatment described in Ref.~\cite{BakerCousins} as an index of the 
degree of consistency between the same- and mixed-event 
invariant-mass distributions in the mass region used for normalization.
We typically found $\chi^2/\mathrm{dof} \approx 0.8$ (for most fits, 
the number of degrees of freedom---one minus the number of populated bins 
outside of the signal window in the mixed-event distributions---was 149).
In the worst case, we found $\chi^2/\mathrm{dof} = 177/149$.

The statistical error on $S$ was calculated as
$\sigma_S^2 = (S + B) + \sigma_B^2$. Here, $(S + B)$ is the contribution from 
Poisson fluctuations in the number of total counts in the signal window, 
and $\sigma_B$ was taken to be $B/\sqrt{B_\mathrm{out}}$,
with $B_\mathrm{out}$ the number of counts in the spectra \emph{outside}
the signal window (where same- and mixed-event spectra are normalized to
the same integral). As an index of the statistics of the measurement,
the values obtained for $S$ range from a few counts for bins in phase space
near the limits of the spectrometer acceptance, to approximately 100 counts 
for bins well within the acceptance. Typical values of the signal-to-total
ratio, $S/(S+B)$, were 0.05--0.30; 74 of the 88 measurements had 
values of $S/(S+B)$ within this interval. The median statistical error
for the 88 measurements of $S$ was 72\%; 57 of the 88 measurements had 
statistical errors of less than 100\%. 

Systematic errors on $S$ may result if the signal window
does not completely include the signal peak. In this case, not only are 
the tails of the signal peak excluded from the signal count---the equal-area
normalization technique will also cause the baseline to be overestimated. 
Systematic errors on $S$ may also arise if the mixed-event distributions 
do not accurately represent the background in the same-event distributions.
Of particular concern is the excess of counts in the mass interval 
1.000--1.010~\GeVcc\ noted in Sec.~\ref{sec:minv} and visible 
in Fig.~\ref{fig:minv}. In the same-event spectra for the individual
centrality and phase-space bins, this excess is present to varying degrees.
Since we are uncertain of the origin of this excess, its effects 
must be included in our estimates of the systematic error for
the signal count in each bin.

\begin{figure}
\includegraphics[width=0.45\textwidth]{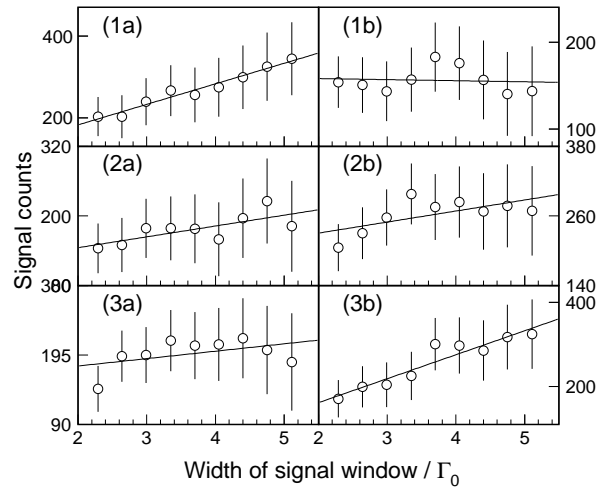}
\caption
{Dependence of $S$ on width of signal window for six bins in $y$ and $m_t$:
1a) $1.2 < y < 1.4$, $1.5 < m_t < 1.7$~\GeVcc,
2a) $1.2 < y < 1.4$, $1.7 < m_t < 1.9$~\GeVcc,
3a) $1.4 < y < 1.6$, $1.7 < m_t < 1.9$~\GeVcc,
1b) $1.2 < y < 1.4$, $1.3 < m_t < 1.5$~\GeVcc,
2b) $1.4 < y < 1.6$, $1.3 < m_t < 1.5$~\GeVcc, and
3b) $1.4 < y < 1.6$, $1.5 < m_t < 1.7$~\GeVcc.
For panels 1a, 2a, and 3a, the spectrometer was at $19^\circ$; for
panels 1b, 2b, and 3b, the spectrometer was at $14^\circ$.
Minimum-bias data (0--77\% central) are shown in each case.
The abscissa is the half-width of the signal window expressed in 
units of the $\phi$ width, $\Gamma_0 = 4.26~\MeVcc$. 
The errors plotted are statistical only.
The lines represent the results of linear, least-squares fits
to the points, using uniform weights, and are used to obtain the 
estimate of the systematic error on $S$ for each bin.}
\label{fig:sbr}
\end{figure}

We assign the systematic error on $S$ for each bin by varying the width 
of the signal window from $\pm2\Gamma_0$ to $\pm5\Gamma_0$.
For window sizes smaller than $\pm2\Gamma_0$, the signal count drops
precipitously. For window sizes larger than $\pm5\Gamma_0$, there is 
very little constraint on the normalization from the low-mass portion 
of the distribution. In Fig.~\ref{fig:sbr} we show the dependence of $S$ 
on the width of the signal window for minimum-bias distributions 
in six bins of $y$ and $m_t$. For each bin, values of $S$ are 
plotted for nine different choices for the signal window, each symmetric 
about the expected peak position with half-width in units of $\Gamma_0$
given by the abscissa value.
The central point in each plot, corresponding to a window
width of about $\pm3.7\Gamma_0$, provides the value of $S$ used 
to obtain the results presented in this work. 
Some of the point-to-point fluctuation
is statistical in nature and is presumably accounted for in the statistical
errors associated with each value of $S$. In most cases, however, there
is a suggestion of an underlying systematic trend. To quantify this trend, 
we perform a linear, least-squares fit with uniform weights to the nine
points for each bin. The results are shown as the straight 
lines in the figure. For each bin, we quote as the systematic error on $S$
the RMS variation of this line over the interval spanned by the
first and last points, which is equal to the maximum vertical extent
divided by $\sqrt{12}$. When applied to each centrality and phase-space
bin used for the analysis, this procedure leads to systematic errors 
for $S$ that vary significantly from bin to bin. For the 88 individual 
measurements of $S$, the median systematic error is 16\%; 62 of the 88
measurements have systematic errors of less than 30\%.
In all cases, the statistical errors on $S$ are larger than the systematic
errors.

The differential yields for each bin in phase space and centrality
were then calculated from $S$ as
\begin{equation}
\frac{1}{m_t}\cdot\frac{d^2\!N}{dy\,dm_t} = 
\frac{1}{\Delta y \Delta m_t}\cdot
\frac{S}{\fZCAL\NINT}\cdot
\frac{\mean{a}\mean{w}}{\mathrm{BR}\mean{m_t}}
\label{eq:counts}
\end{equation}
where 
\fZCAL\ is the fraction of minimum-bias interaction triggers satisfying the 
centrality selection,
\NINT\ is the background-subtracted number of minimum-bias interaction 
triggers collected with the data for normalization purposes,
\mean{a} is the geometrical acceptance correction,
\mean{w} is the overall correction factor for other experimental effects,
\mean{m_t} is the mean value of $m_t$ for all pairs in the bin, and
$\mathrm{BR}(\Pphi\rightarrow\PKp\PKm) = 49.2$\%.

Apart from the systematic error on $S$, the principal source of systematic
uncertainty in the yields is from the uncertainty on \fZCAL.  
Due to radiation damage to the plastic-scintillator materials
of the ZCAL, its signal decreased with time.
A run-dependent calibration was therefore used, in which the energy of the 
fragmented-beam peak was recentered at the nominal beam kinetic energy
(with corrections for energy loss in the target). 
These calibrations were performed periodically throughout the 
running period.
The centrality bins were then defined via fixed cuts on the 
calibrated ZCAL energy.
However, \fZCAL\ exhibited residual variation as a function 
of running time.
The resulting contributions to the systematic errors on the yield 
measurements are estimated to range from about $5\%$ in
the most central bin to $10\%$ for the most peripheral bin.

The geometrical acceptance, \mean{a}, and many of the contributions to
the weight factor \mean{w}, were calculated using a GEANT-based Monte
Carlo simulation implementing a realistic detector
configuration~\cite{E917PHD}.  The value of \mean{a} used in
\Eq{eq:counts} is the average value for $\phi\rightarrow K^+K^-$
events in the phase-space bin of interest.  The weight factor \mean{w}
was calculated as follows for each centrality and phase-space bin.
Corrections for kaon decays in flight, the single-particle tracking
inefficiency, hadronic interactions, multiple scattering in the
spectrometer, momentum cuts, and the kaon-identification inefficiency
were evaluated using the Monte Carlo, separately for each track in
each pair.  An additional correction for the single-track tracking
efficiency in the presence of background hits on the drift chambers
was evaluated by inserting found tracks into random events. This
occupancy correction was then parameterized by the hit multiplicity in
the chambers and applied for each track; the resulting correction was
typically 20--40\%.  The individual correction weights for each track
and from each of the above sources were then multiplied, together with
an additional correction for a two-track opening-angle cut (evaluated
for each pair), to obtain the overall pair weight, $w$.  The value of
\mean{w} in \Eq{eq:counts} is the average value of $w$ for all pairs
in the same-event distribution of interest. The systematic errors
associated with \mean{a} and \mean{w} are estimated to be no larger
than 5\%, arise mainly from the statistics of the samples used to
obtain the weight factors, and do not significantly affect the
results.

The final systematic errors on the individual yield measurements are
calculated as the quadrature sum of the fractional systematic errors
on $S$ (for each measurement), on \fZCAL (for each centrality bin), 
and on \mean{a}\mean{w} (constant).
In general, the dominant contribution is from the systematic uncertainty
on $S$. For the 88 individual differential yield measurements,
the median overall systematic error is 20\%; 60 of the 88
measurements have systematic errors of less than 30\%, 
and 75 of the 88 have systematic errors of less than 50\%. 
In all cases, the statistical errors are larger than the overall systematic
errors; for 73 of the 88 measurements, the statistical errors are more 
than twice as large as the overall systematic errors.

Experimental results from both the $14^\circ$ and $19^\circ$ settings
of the spectrometer were in all cases consistent within statistical errors,
and were therefore combined for the presentation of the transverse-mass
spectra, resulting in the 53 individual differential yield measurements 
plotted in Fig.~\ref{fig:mtspec}.

\subsection{Transverse-mass spectra and rapidity distributions}

\begin{figure}
\includegraphics[width=0.48\textwidth]{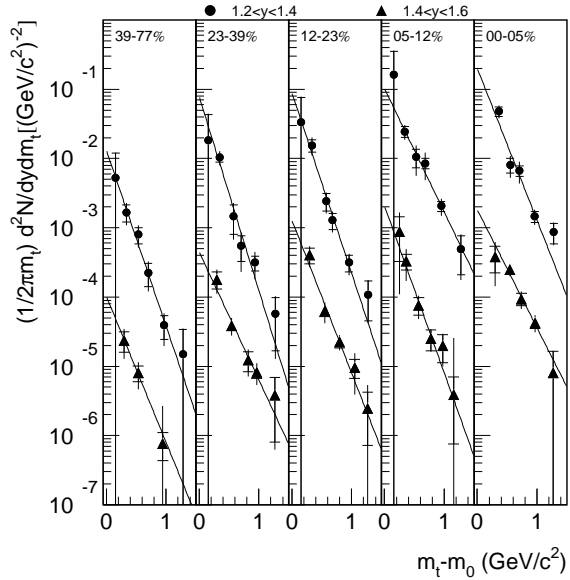}
\caption
{The invariant yield of \Pphi's as a function of transverse mass
($m_t-m_0$) in five centrality bins labeled by $\sigma/\sigint$.  In
each panel, the top set of points (circles) is for $1.2<y<1.4$, while
the second set of points (triangles) is for $1.4<y<1.6$ and is
divided by 100 for presentation. The total errors are calculated as
the quadrature sum of the systematic and statistical errors, while the
magnitudes of the latter are indicated by the cross bars. The lines are
the results of the exponential fits described in the text.}
\label{fig:mtspec}
\end{figure}

The transverse-mass spectra for the five centrality bins are shown in
two bins of rapidity, $1.2 < y < 1.4$ and $1.4 <y < 1.6$, with an
$m_t$ bin size of 0.2~\GeVcc, in Fig.~\ref{fig:mtspec}. For the $m_t$
bins where the acceptance from $14^\circ$ and $19^\circ$ data sets
overlap, the individual measurements from each of the data sets agree 
to within statistical errors, and the weighted average of the two
measurements is presented. The full error shown includes both
statistical and systematic contributions; the contribution from
statistics alone is indicated by the cross bars. The total errors are
calculated as the quadrature sum of the statistical and systematic
errors.

An exponential parameterization with two parameters, rapidity density
(\dndy), and inverse slope ($T$), was used to fit the transverse-mass
spectra:

\begin{eqnarray} 
\frac{1}{2 \pi c^4 m_t} \frac{d^2N}{dm_t dy} &=&
\frac{\dndy}{2\pi (T m_0 c^2+ T^{2})} \nonumber \\
& & \times \exp\left[-\frac{(m_t-m_0)c^2}{T}\right]. 
\end{eqnarray} 

\begin{figure}
\includegraphics[width=0.45\textwidth]{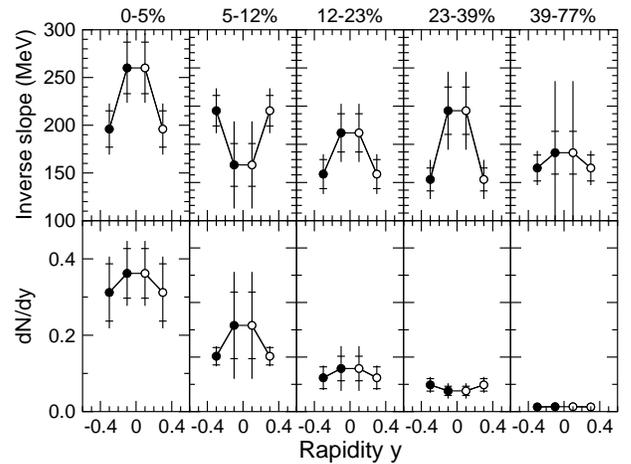}
\caption
{The inverse-slope parameters, $T$, and rapidity density, \dndy, of
\Pphi\ mesons as a function of centrality. The open symbols show the
reflections of the data points about midrapidity ($y=1.6$). The error 
bars have the same significance as in Fig.~\ref{fig:mtspec}.}
\label{fig:dndy}
\end{figure}

This function gave a good fit to the data with $\chi^2/\mathrm{dof}
\about{1}$ in all cases. The values obtained for the rapidity density (\dndy)
and inverse-slope parameters ($T$) are plotted as functions of
rapidity and centrality in Fig.~\ref{fig:dndy} and are listed in
Tables~\ref{tab:tinv} and~\ref{tab:dndy}.
The statistical and systematic contributions to the error were evaluated
by performing the fits with and without the systematic contribution included
in the errors. The systematic error due to uncertainties in the overall 
normalization is excluded from the systematic error estimate for $T$.

As a function of centrality,
the rapidity density shows a strong systematic increase, while the
inverse slope increases more mildly. Within the rapidity range covered
by the measurement, there seems to be no strong rapidity dependence
for $T$.

\begin{figure}
\includegraphics[width=0.45\textwidth]{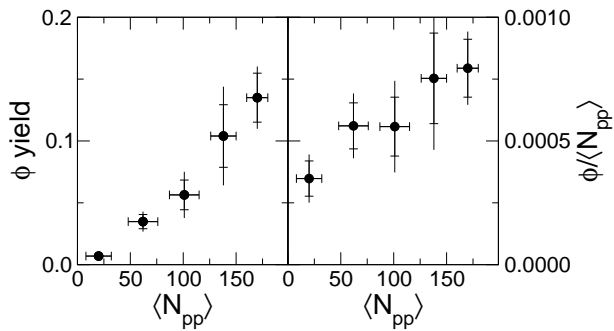}
\caption
{The fiducial \Pphi\ yield (left) and fiducial \Pphi\ yield per
projectile participant, \mean{\npp}, as a function of \mean{\npp} (right).
The accepted rapidity region is $1.2<y<1.6$.  The error bars have the
same significance as in Fig.~\ref{fig:mtspec}.}
\label{fig:ratio_phinpp}
\end{figure}

The fiducial yield is the sum of the values $\dndy \times \Delta y$
for the two rapidity bins covering the interval $1.2 < y < 1.6$ and is
plotted as a function of \mean{\npp} (calculated from a Glauber model
as explained in Sec.~\ref{sec:expt}) in Fig.~\ref{fig:ratio_phinpp}.
When normalized to \mean{\npp}, this quantity exhibits a steady
increase with increasing centrality. This implies that the fiducial yield of
\Pphi's increases faster than linearly with \mean{\npp}. It is
noted that the increase with centrality is slightly more 
significant (the effect is on the order of 4\% in the slope of a linear fit) 
if the fiducial yields are plotted against \mean{\nppZCAL}. Thus, we observe
an enhancement in \Pphi-meson production in central events. This is
similar to what has been previously observed for kaon
production~\cite{E866KAON}. In \coll{Pb}{Pb} collisions at the SPS, 
NA50 has also observed an increase in the yield of \Pphi's per participant
nucleon (\mean{n_\mathrm{p}}) as a function of 
\mean{n_\mathrm{p}}~\cite{NA50PHI}, 
though the NA50 data show more evidence of saturation of this quantity 
in the most central collisions than do our results.

\begin{table}
\caption
{Inverse-slope parameter, $T$, in units of \MeVcc, for \Pphi's in two
bins of rapidity and five bins of centrality. Errors are statistical
followed by systematic.}
\begin{ruledtabular}
\begin{tabular*}{0.50\textwidth}{c c c}
Centrality & $1.2 < y < 1.4$ & $1.4 < y < 1.6$ \\\hline
0--5\%        & $  196 \pm  19 \pm  19$ & $  260 \pm  27 \pm  25$ \\
5--12\%       & $  244 \pm  20 \pm  22$ & $  173 \pm  28 \pm  50$ \\
12--23\%      & $  161 \pm  18 \pm  19$ & $  215 \pm  28 \pm  25$ \\
23--39\%      & $  154 \pm  15 \pm  21$ & $  244 \pm  31 \pm  40$ \\
39--77\%      & $  169 \pm  17 \pm  15$ & $  189 \pm  28 \pm  90$ \\
\end{tabular*}
\end{ruledtabular}
\label{tab:tinv}
\end{table}

\begin{table}
\caption
{Rapidity density, \dndy, for \Pphi's in two rapidity bins, by
centrality. Errors are statistical followed by systematic.}
\begin{ruledtabular}
\begin{tabular*}{0.50\textwidth}{c c c}
Centrality & $1.2 < y < 1.4$ & $1.4 < y < 1.6$ \\\hline
0--5\%       & $ 0.312 \pm 0.075 \pm 0.056$ & $ 0.362 \pm 0.065 \pm 0.054$ \\
5--12\%      & $ 0.203 \pm 0.032 \pm 0.018$ & $ 0.316 \pm 0.122 \pm 0.153$ \\
12--23\%     & $ 0.124 \pm 0.040 \pm 0.020$ & $ 0.158 \pm 0.045 \pm 0.069$ \\
23--39\%     & $ 0.098 \pm 0.023 \pm 0.018$ & $ 0.076 \pm 0.017 \pm 0.022$ \\
39--77\%     & $ 0.017 \pm 0.005 \pm 0.004$ & $ 0.018 \pm 0.005 \pm 0.005$ \\
\end{tabular*}
\end{ruledtabular}
\label{tab:dndy}
\end{table}

\section{Discussion}
\label{sec:discussion}

In this section, we compare the measured inverse slopes ($T$) and
rapidity densities ($dN/dy$) for the \Pphi\ to those of the other 
hadrons and explore their dependence on centrality and beam energy.

\subsection{Transverse-mass spectrum of the \Pphi; comparison to other species}
\label{sec:disc_tmass}

In the most central bin ($\sigma/\sigint = \mbox{0--5}\%$), the
inverse-slope parameter is determined to be 260$\pm$37~\MeVcc\ around
midrapidity.  As motivated by hydrodynamic models~\cite{hydroPRC}, the
inverse-slope parameters for different particle species are expected
to scale with species mass for particles that participate in the
collective transverse flow. It is interesting to see whether the
inverse slope for \Pphi\ mesons fits into the systematic trend
observed for the other hadrons. However, two known effects complicate
the interpretation.

The first is that the cross sections for \Pphi\ interactions in the
collision medium are expected to be small. The OZI rule inhibits 
elastic scattering of the \Pphi\ on nucleons and non-strange mesons and 
resonances. Elastic-scattering cross-sections are dominated by Pomeron 
exchange and are expected to be on the order of 1~\mb\ (see \cite{SHOR}
and references therein).
In particular, from studies of \Pphi\ photoproduction, 
total cross sections for the interaction of \Pphi's with nucleons 
are estimated to be 8--12~\mb\ over the range of \Pphi\ 
kinetic energies from a few hundred \MeV\ up to several \GeV, and 
dominated by OZI-allowed absorptive processes---elastic scattering 
of \Pphi's on nucleons accounts for about 1~\mb\ of this cross
section~\cite{PHICROSS}.
It might therefore be expected that \Pphi's decouple rather early from the
nuclear fireball, which is composed of nucleon resonances, pions, 
and non-strange meson resonances. If this were 
the dominant effect, the \Pphi\ inverse slope ($T$) would
reflect the original
``temperature'' when the \Pphi's are produced, without the enhancement
from collective transverse flow developed in the late hadronic
stages~\cite{Xu_PRL}. In this case, $T$ for the \Pphi's would be lower
than that observed for other hadrons of similar mass (for example,
$p$'s).

On the other hand, when the \Pphi\ is detected via its decay into
kaons, the rescattering of the daughter kaons from \Pphi's that decay
inside the nuclear fireball would generate a relative depletion in the
\Pphi\ yield at low $p_t$.
This explanation has been proposed~\cite{SHURYAKQM99} to explain
the difference in the values of $T$
for \Pphi\ mesons obtained by experiments NA49 and NA50, which observed
the \Pphi\ via its decay into \PKp\PKm\ and $\mu^+\mu^-$ final states,
respectively.
Given the short lifetime of the \Pphi\
($c\tau = 45$~fm/$c$), one might expect that this effect can give 
rise to a significant depletion in the $m_{t}$ spectrum only at low $m_t$. 
To illustrate, we have performed a simple simulation in which \Pphi's
are generated at the center of a ``fireball'' according to an exponential
$m_t$ distribution with $T=250~\MeVcc$, and any \Pphi's that decay within
7~fm of the origin are suppressed.
For $m_t - m_0 > 0.5~\GeVcc$ the resulting depression of the $m_t$
spectrum is nearly uniform; a deficit in the spectrum of more than 
10\% is noted only for $m_t - m_0 \lesssim 0.25~\GeVcc$, \emph{i.e.}, 
only at the edge of the E917 acceptance. 
(A calculation performed with RQMD for \coll{Pb}{Pb} collisions at the 
SPS gives similar results~\cite{JJD}.)
Conceivably, however, in-medium effects could broaden the \Pphi\ enough
to cause a more significant fraction of \Pphi's to decay inside the fireball,
leading to an increase in the apparent value of $T$ in fits to the 
$m_t$ spectra (see Ref.~\cite{KF} for an example with reference to the 
NA49 and NA50 data).

For the comparison of the transverse-mass spectra for various particle
species, we choose to use the mean transverse mass, $\mean{m_t}$.
This is because a variety of forms are used to fit and parameterize
the transverse-mass spectra in the existing measurements of 
$\pi^+$~\cite{E866PIPRO}, 
$K^+$~\cite{E866KAON},
$p$~\cite{E866PROD,E917PROC},
$d$~\cite{E866PROD},
and $\bar{\Lambda}$~\cite{E917LBAR} production at midrapidity
in central \coll{Au}{Au}
collisions at the AGS, including
Boltzmann, Boltzmann plus exponential, and $m_t$-scaled exponential forms, 
in addition to the simple single-exponential form used to fit
the $m_t$ spectra for $\phi$'s in this work.
In all but two of the reports referenced above, $\mean{m_t}$ is derived from 
the slope parameters for the chosen parameterization of the $m_t$
spectra, and quoted in place of the slope parameters themselves.
The exceptions are the $K^+$ and $\bar{\Lambda}$ data.
In Ref.~\cite{E866KAON}, the $m_t$ spectra for kaons are fit with 
exponential distributions, and values of $T$ are reported. We convert 
to $\mean{m_t}$ using the expression
\begin{equation}
\mean{\mbox{$m_t$}}_\mathrm{exp} = 
\frac{m_0^2 c^4 + 2 m_0 c^2 T + 2T^2}{m_0 c^2 + T}.
\label{eq:Texp}
\end{equation}
We use the same expression to obtain $\mean{m_t}$ from the value 
of $T$ obtained from the fit to the $m_t$ spectrum for the $\phi$
presented in this work.
In Ref.~\cite{E917LBAR}, the $m_t$ spectra for $\bar{\Lambda}$'s
are fit with Boltzmann distributions,
\begin{eqnarray} 
\frac{1}{2 \pi c^4 m_t} \frac{d^2N}{dm_t dy} & = &
\frac{\dndy}{2\pi (\TB
m_0^2 c^4 + 2 \TB^2 m_0 c^2 + 2 \TB^{3})} \nonumber \\ 
& & \times m_t c^2
\exp\left[-\frac{(m_t - m_0) c^2}{\TB}\right],
\end{eqnarray}
and values for $\TB$ are quoted. We convert 
to $\mean{m_t}$ using the expression
\begin{equation}
\mean{\mbox{$m_t$}}_\mathrm{Boltz}  =  
\frac{m_0^3 c^6 + 3m_0^2 c^4 \TB + 6 m_0 c^2 \TB^2 + 6 \TB^3}
{m_0^2 c^4 + 2 m_0 c^2 \TB + 2 \TB^2}.
\label{eq:TBoltz}
\end{equation}

\begin{figure}
\includegraphics[width=0.48\textwidth]{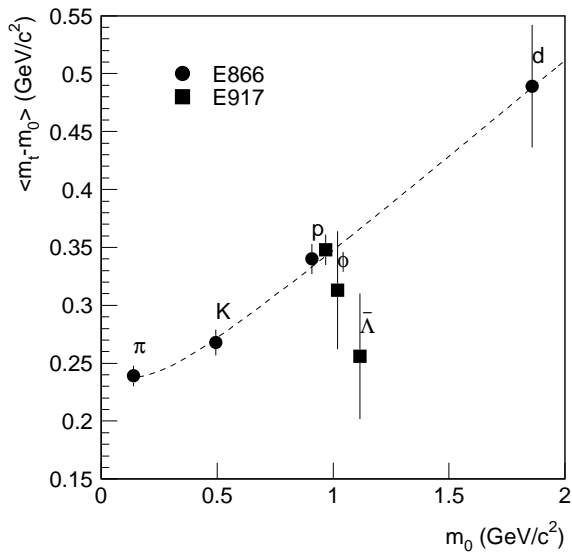}
\caption
{The mean transverse mass (\mean{m_t-m_0}) for various particle
species as a function of the species mass at midrapidity for central
\coll{Au}{Au} collisions at 11.7~\GeVc\ per nucleon as measured by E866 and
E917. The dashed line is a fit assuming exponential transverse-mass spectra
and a linear relationship between inverse slope ($T$) and particle
mass ($m_0$), as described in the text. Values of $T$ for particles other
than the \Pphi\ are taken from
Refs.~\cite{E866KAON,E917LBAR,E917PROC,E866PIPRO,E866PROD}. 
The proton data point
from E866 (E917) is presented with a shift of $-30$~\MeVcc\
($+30$~\MeVcc) in mass for clarity.}
\label{fig:tmassvsmass}
\end{figure}

The values of \mean{m_t-m_0} for \Pphi\ mesons, together with the
corresponding values for $\pi^+$'s, $K^+$'s, $p$'s, $d$'s, and
$\bar{\Lambda}$'s are plotted as a function of species mass in
Fig.~\ref{fig:tmassvsmass}. We note that the centrality selection
differs somewhat from species to species in the available data: top
$3\%$ for $\pi^+$, $p$ (E866), and $d$; top $5\%$ for $p$ (E917), \Pphi\
and $K^+$; and top $12\%$ for $\bar{\Lambda}$. 
The rapidity coverage varies similarly: 
for the $\pi^+$ data it is $1.3<y<1.4$; 
for the $K^+$ and $\phi$ data it is $1.4<y<1.6$, 
for the $p$ and $d$ data it is $1.4<y<1.5$,
and for the $\bar{\Lambda}$ data it is $1.0<y<1.4$.
(For reference, $y_{NN} = 1.61$ for 11.7\AGeVc\ \coll{Au}{Au} collisions.)
The dashed line in Fig.~\ref{fig:tmassvsmass} shows a fit to the values 
for $\mean{m_t-m_0}$ using the form of \Eq{eq:Texp} with $T=a+bm_0c^2$, 
so that $a$ and $b$ are the free parameters of the fit. 
Assuming that it is qualitatively valid to describe
the midrapidity $m_t$ spectra for the various species
by exponential inverse-slope parameters, this fit 
illustrates the trend expected for a linear relationship between
inverse slope ($T$) and particle mass ($m_0$), as motivated 
by hydrodynamic models.

The value of \mean{m_t-m_0} for the \Pphi\ falls slightly below the
systematic trend observed for the other particles, but less so than
for $\bar{\Lambda}$'s. There may be some suggestion of an effect
from early freeze-out of the \Pphi\ and late development of at least 
some of the transverse flow, as described in Ref.~\cite{Xu_PRL}.
However, the significance of this observation is limited by the 
precision of the measurement.

\subsection{Centrality dependence of \Pphi\ production; comparison to
other species}
\label{sec:disc_cent}

As mentioned in Sec.~\ref{sec:intro}, processes such as $pp
\rightarrow pp \phi$ are suppressed by large threshold energies
and the OZI effect. Thus a naive expectation assuming ordinary
hadronic interactions is that if an enhancement in \Pphi\ production
were observed in heavy-ion collisions at AGS energies, this
enhancement would result from secondary collisions, \emph{e.g.} via
channels such as $\PKp \Lambda \rightarrow \phi p $ or $\PKp \PKm
\rightarrow \Pphi \rho$~\cite{ART3}. If this were in fact the
case, increasing $KY$ and $K \bar{K}$ combinatorics would bring about
an increase of the $\Pphi/K$ ratio for central collisions.

On the other hand, the proposed ``re-arrangement''~\cite{PARTONFUSION}
or ``shake-out''~\cite{KNOCKOUT} of an intrinsic $s\bar{s}$ component
of the nucleon wave function in the non-perturbative regime provides a
mechanism for \Pphi\ production in \coll{N}{N} interactions that
bypasses the effects of OZI suppression. 
If the $s\bar{s}$ component of the nucleon wave function were
negatively polarized, the fact that, at threshold,
the reaction $p p \rightarrow p p \Pphi$ must proceed with the $pp$
initially in the $^3S_1$ spin state would imply that 
this mechanism should be particularly important at near-threshold energies
(see Ref.~\cite{DISTO} and references therein, especially~\cite{ELLIS2}). 
Indeed, the DISTO collaboration has observed that in \ppcoll\ collisions
at $\sqrt{s} = 2.90~\GeV$, only 83~\MeV\ above threshold, 
the $\Pphi/\PKm$ ratio is about unity, such that \Pphi\ production
represents an important contribution to the \PKm\ yield at these 
energies~\cite{DISTO}.

Along these lines, a possible mechanism has been proposed to explain
overall strangeness production in \coll{N}{A} collisions within the
framework of the additive quark model~\cite{ADM}. According to this
proposal, strange particles are born as strange-quark pairs from
binary collisions of the projectile and target nucleons, with a
probability proportional to the number of interacting constituent
projectile quarks~\cite{KADIJA}. In \pAcoll{Au} collisions at
17.5~\GeVc, E910 has shown that the production of $\Lambda$'s and
\PKS's increases with the estimated number of binary collisions,
$\nu$, suffered by the incident proton~\cite{E910PRL}. For $\nu \le
3$, the increase in the yields is faster than expected from scaling of
\ppcoll\ data by the number of total participants (\emph{i.e.},
$N_{p+\mathrm{A}}/N_{p+p} = \frac{1}{2}(1 + \nu)$), although bounded
from above by linear scaling with $\nu$ (\emph{i.e.},
$N_{p+\mathrm{A}}/N_{p+p} = \nu$).

If the same mechanism were responsible for the observed increase in
\Pphi\ production with centrality in \coll{A}{A} collisions, we would
expect to observe similar scaling behavior in our data. Specifically,
we would expect approximately constant numbers of hadrons bearing
strange quarks to be produced per binary collision. In the following,
we therefore compare the yield of \Pphi's to the yields of pions and
kaons as a function of centrality. The yields of pions and kaons were
obtained by the E866 collaboration from \coll{Au}{Au} collisions at
the same beam energy used for the present
measurement~\cite{E866KAON,E866PIPRO}.

\begin{figure}
\includegraphics[width=0.45\textwidth]{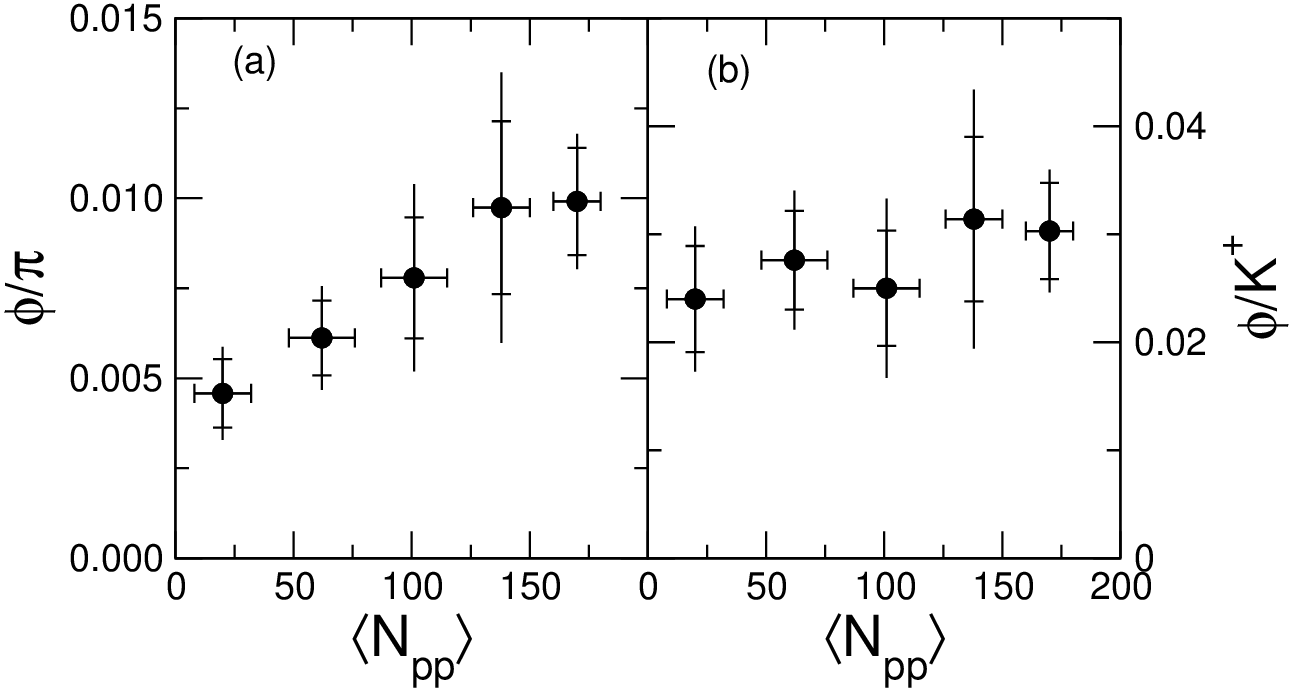}
\caption
{The $\phi/\pi$ and $\Pphi/\PKp$ ratios as a function of \mean{\npp}.
  For the \Pphi, the fiducial yields from E917 in the rapidity
  interval $1.2 < y <1.6$ are used.  For $\pi$'s, the data have been
  taken from Ref.~\cite{E866PIPRO}, and represent fiducial yields in
  the rapidity interval $1.2 < y < 1.4$.  The value used for the $\pi$
  yield is 1.125 times the yield for $\pi^+$, which is essentially the
  average yield for $\pi^+$ and $\pi^-$~\cite{E866PROPI}.  For \PKp's,
  the data have been taken from Ref.~\cite{E866KAON}, and represent
  fiducial yields in the rapidity interval $1.2 < y < 1.6$.}
\label{fig:ratio_phi}
\end{figure}

As seen from the left panel of Fig.~\ref{fig:ratio_phi}, the
$\Pphi/\pi$ ratio of fiducial yields shows a rise toward central
collisions, which signals an enhancement in the production of \Pphi\
mesons relative to that of non-strange $\pi$ mesons in central
collisions. This enhancement is also clearly suggested by
Fig.~\ref{fig:ratio_phinpp}, since pion production is known to mainly
come from resonance decay and secondary rescattering and scales
linearly with \mean{\npp}. Due to the fact that the plotted points
represent ratios of fiducial yields, some care should be taken when
interpreting the exact form of the dependence.  We note that this
increase in the $\Pphi/\pi$ ratio is qualitatively similar to the
increase in the $K/\pi$ ratio with centrality observed by E866 in
\coll{Au}{Au} collisions at the AGS~\cite{E866PIPRO}.

We next compare the degree of enhancement in the yields of \Pphi's and
$K$'s, which both contain $s$ quarks.  We use the yield of \PKp\ to
characterize the overall kaon production because the E866 data
indicate that the $\PKm/\PKp$ ratio is about 0.15 and independent of
centrality. In the right plot of Fig.~\ref{fig:ratio_phi}, the ratio
of fiducial yields $\Pphi /\PKp$ is plotted as a function of
centrality, and shows no substantial variation. This implies that both
\Pphi\ and \PKp\ (or \PKm) possess a similar degree of enhanced
production toward central collisions at AGS energies. 
As estimated from linear fits to the $\Pphi /\PKp$ ratios as a function 
of \mean{\npp}, an increase of up to 50\% for the ratio in the most
central bin cannot be definitively ruled out within a $1\sigma$ range
of the fit errors. However, any centrality dependence of the $\Pphi/K$ ratios
appears to be weak.

\begin{figure}
\includegraphics[width=0.48\textwidth]{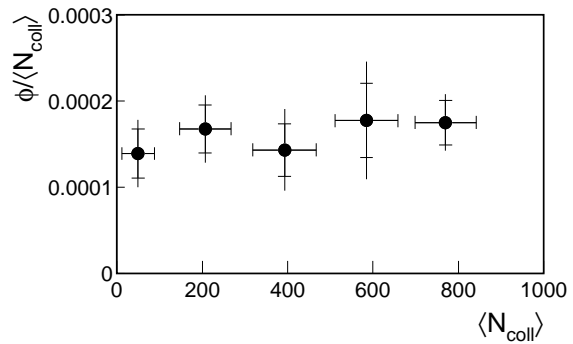}
\caption
{The normalized fiducial yield of \Pphi\ mesons per number of binary
collisions, \mean{\ncoll}, as a function of \mean{\ncoll}. The
accepted rapidity region is $1.2<y<1.6$.}
\label{fig:ratio_phincoll}
\end{figure}

To look for scaling behavior similar to that observed in \pAcoll{Au}
collisions by E910, we examine the dependence of the \Pphi\ yields in
our data on the mean number of binary collisions,
\mean{\ncoll}. (Estimation of \mean{\ncoll} is discussed in
Sec.~\ref{sec:expt}; the values of \mean{\ncoll} for each centrality
bin are given in Table~\ref{tab:cent}.) The fiducial \Pphi\ yield
normalized to \mean{\ncoll} is plotted versus \mean{\ncoll} in
Fig.~\ref{fig:ratio_phincoll}. The dependence of the fiducial \Pphi\ yield
on \mean{\ncoll} is consistent with scaling proportional to \mean{\ncoll}.

While the rapidity coverage of our measurement does not allow us to
make a precise statement about the absolute value of the \Pphi\ yield,
it is possible to make an informed guess about the width of the \Pphi\
rapidity distribution. Both E859~\cite{E859PRL} and
NA49~\cite{NA49PHI,NA49KPI} have observed that the Gaussian width of
the \Pphi\ rapidity distribution is very similar to those of the \PKp\
and \PKm\ rapidity distributions. In central \coll{Au}{Au} collisions
at the AGS, the Gaussian widths of the rapidity distributions for
\PKp\ and \PKm\ are \about{0.9} and \about{0.7},
respectively~\cite{E866KAON,E859KAA}. The narrow \PKm\ width has been
attributed to the higher threshold and more restricted phase space for
\PKm\ production. Since the thresholds for \PKm\ and \Pphi\ production
in \ppcoll\ collisions are very similar, we assume that the \Pphi\ and
\PKm\ have Gaussian rapidity distributions with approximately equal
widths in central \coll{Au}{Au} collisions. We then estimate that our
fiducial yield corresponds to about 20\% of the total yield. Using the
parameterization described in Appendix~\ref{appendix} for the \Pphi\
yield in \ppcoll\ collisions as a function of center of mass energy,
$\sqrt{s}$, the yield of \Pphi's per binary collision in \coll{Au}{Au}
collisions is about 50\% of the \Pphi\ yield in \ppcoll\ collisions at
this energy. This observation seems contrary to the usual expectation
of enhanced strangeness production in \coll{Au}{Au} collisions. We
discuss this point further in Sec.~\ref{sec:disc_excit}.

Using E866 data on the total yields of \PKp\ and \PKm\ normalized to
\mean{\ncoll}, a similar scaling proportional to \mean{\ncoll} is
observed. It appears as if the mechanisms for the
production of \PKp's, \PKm's, and \Pphi's all have a similar
dependence on the centrality of the collisions, and this dependence is
consistent with a scaling with the number of binary collisions,
\mean{\ncoll}. This observation suggests that hard binary
collisions might play an important role in strangeness production
in heavy-ion collisions.

Our observation that \Pphi\ and kaon production scale similarly and
faster than linearly with \mean{\npp}\ therefore provides an essential
test of the details contained within rescattering models. The
rescattering model that suggests the importance of contributions
from $KY$ and $K \bar{K}$ collisions in the increase of \Pphi\
production in central \coll{A}{A} collisions~\cite{ART3} can be
ruled out; if such contributions were dominant, combinatorial
considerations would lead to the expectation that \Pphi\ production
should increase faster than kaon production with centrality. This is
incompatible with our observation that \Pphi\ and kaon production
show similar scaling with centrality.

\subsection{Dependence of $\Pphi/\pi$, $\Pphi/\PKp$, and $\Pphi/\PKm$
ratios on \sNN\ in \coll{A}{A} and \ppcoll\ reactions}
\label{sec:disc_excit}

\begin{figure}
\includegraphics[width=0.48\textwidth]{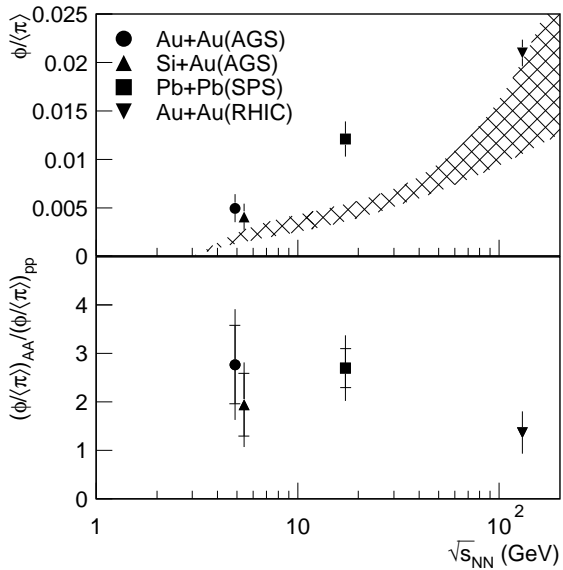}
\caption
{Ratios of $\phi/\mean{\pi}$ in central heavy-ion and \ppcoll\
  interactions as a function of \sNN\ (top panel) and double ratios of
  \coll{A}{A} over \ppcoll\ (bottom panel).  For the E917 and STAR
  points, fiducial yield ratios are plotted; for the other two points,
  total yield ratios are shown instead.  The hatched area represents
  the ratio of total yields in \ppcoll\ collisions, based on the
  parameterization discussed in Appendix~\ref{appendix}. The cross
  bars in the bottom panel indicate the contribution to the total 
  errors from the uncertainty 
  on $\phi/\mean{\pi}$ in central heavy-ion interactions.}
\label{fig:excit_phipi}
\end{figure}

\begin{figure}
\includegraphics[width=0.48\textwidth]{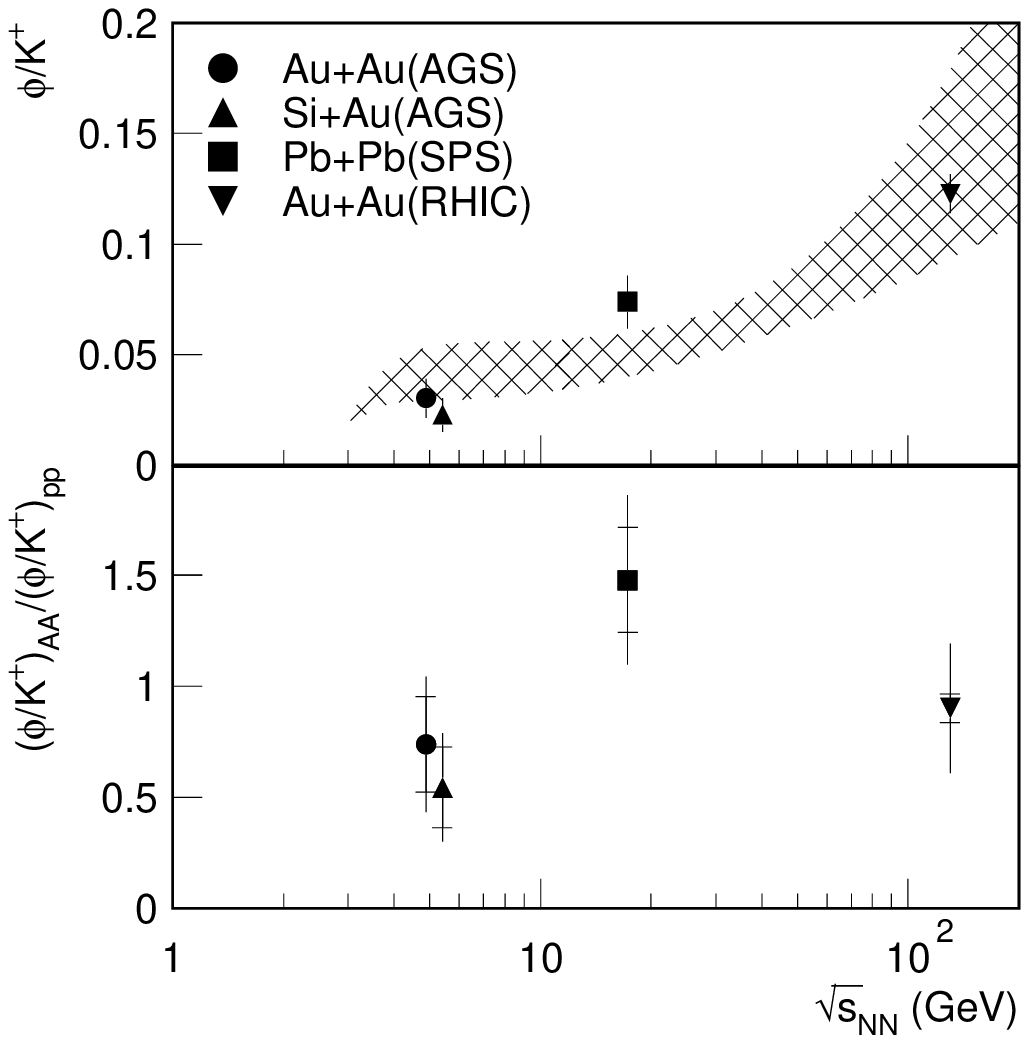}
\caption
{Same as Fig.~\ref{fig:excit_phipi} for the ratios of $\phi/\PKp$.}
\label{fig:excit_phikp}
\end{figure}

\begin{figure}
\includegraphics[width=0.48\textwidth]{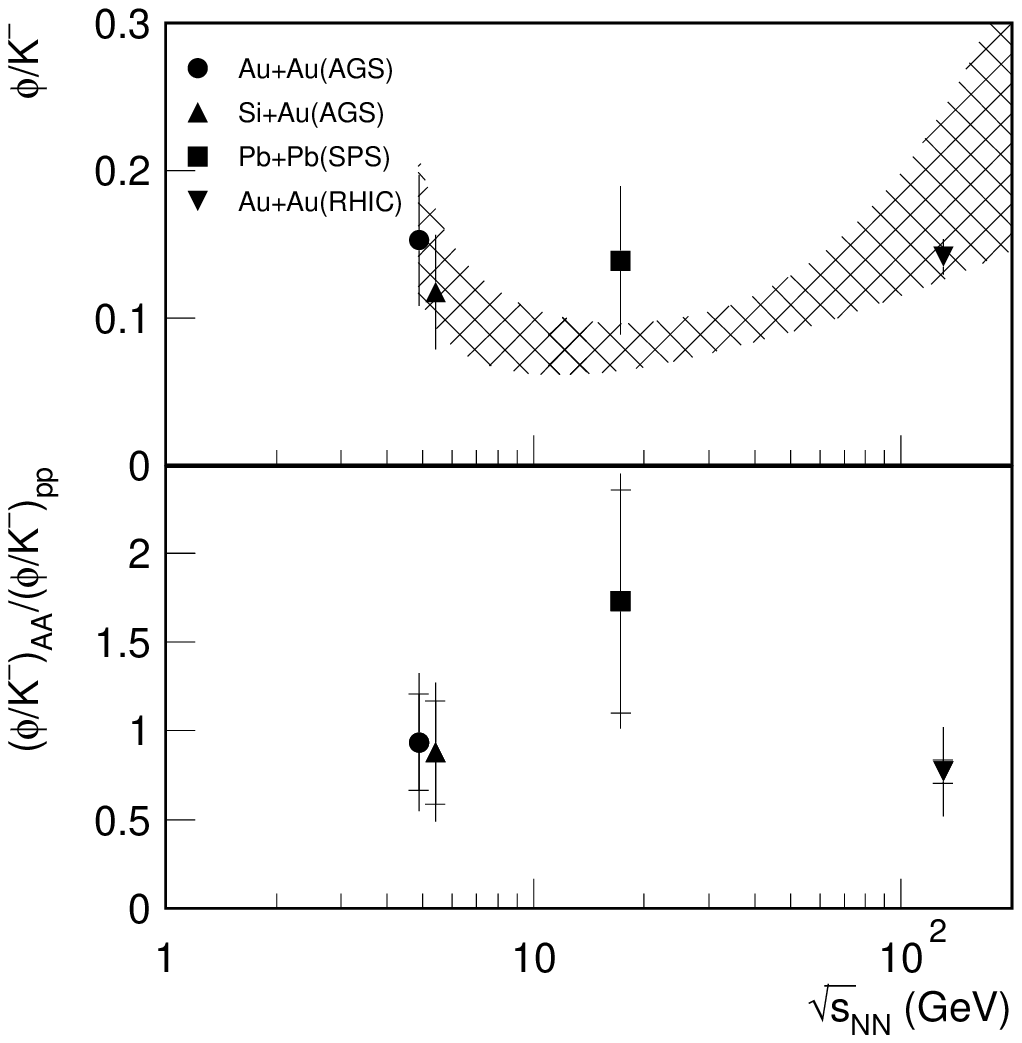}
\caption
{Same as Fig.~\ref{fig:excit_phipi} for the ratios of $\phi/\PKm$.}
\label{fig:excit_phikm}
\end{figure}
 
In order to further explore the mechanism responsible for the
approximate scaling of the \Pphi\ yield with \mean{\ncoll} observed in our
data, we compare our results with other measurements of \Pphi\
production in heavy-ion collisions at different reaction energies.

The excitation functions of the $\phi/\pi$, $\phi/\PKp$, and
$\phi/\PKm$ ratios in central heavy-ion collisions are shown in
Figs.~\ref{fig:excit_phipi},~\ref{fig:excit_phikp},
and~\ref{fig:excit_phikm}. The four points correspond to \coll{Au}{Au}
collisions at the AGS (this measurement, $\sNN =
4.87~\GeV$)~\cite{E866KAON,E866PIPRO}, \coll{Si}{Au} collisions at the
AGS ($\sNN = 5.39~\GeV$)~\cite{E859PRL,E802KPI,E859KAA}, \coll{Pb}{Pb}
collisions at the SPS ($\sNN = 17.27~\GeV$)~\cite{NA49PHI,NA49KPI},
and \coll{Au}{Au} collisions at RHIC ($\sNN =
130~\GeV$)~\cite{STARPHI,PHENIXKPI}.  These plots must be interpreted
with some care.  Most obviously, the collisional system is different
in each case; in particular, the various particle yields are not
guaranteed to scale in the same way when passing from the
\coll{Si}{Au} system to the heavier systems. In addition, the points
for \coll{Au}{Au} collisions at the AGS and RHIC represent
fiducial-yield ratios (the AGS point is for the fiducial yield over
$1.2 < y < 1.6$, and the RHIC point is for the central unit of
rapidity), while the other two points are ratios of yields over all
phase space. Nor have we applied any corrections in the comparisons
with \ppcoll\ collisions to take into account the isospin averaging of
yields from \ppcoll, \nncoll, and \pncoll\
collisions~\cite{E866KAON,E859KAA,ISOSPIN}. Nevertheless, two
observations can be made.

Our first observation is that the $\Pphi/\pi$ ratios for \coll{A}{A}
collisions are notably enhanced with respect to their values for
\ppcoll\ collisions for all points except at the highest energy in
Fig.~\ref{fig:excit_phipi}.  In \ppcoll\ collisions, $\Pphi/\pi$
increases with \sNN, in large part because of the larger production
threshold for \Pphi\ mesons. In heavy-ion collisions, the $\Pphi/\pi$
ratios are enhanced by factors of 2--3, at least up to SPS energies,
but seem to show the same energy dependence observed in \ppcoll\
collisions, at least up to SPS energies.  Such an enhancement in the
central \coll{A}{A} collisions could be interpreted as due to the different
scaling behaviors of \Pphi\ and $\pi$ with centrality. As discussed
in Sec.~\ref{sec:disc_cent}, the yield of \Pphi's scales with
\mean{\ncoll} while that of $\pi$'s scales with \mean{\npp}. The ratio
of \mean{\ncoll} to \mean{\npp} is around 1 in peripheral
collisions, which are similar to \ppcoll\ collisions, and becomes 
larger than 1 in central collisions. At RHIC, the $\Pphi/\pi$ ratio for central
\coll{Au}{Au} collisions is about the same as that from the
parameterization of the \ppcoll\ data. It might be speculated that the
primary production channels for \Pphi's and/or $\pi$'s at RHIC
energies are different from those at AGS and SPS energies.

The second observation is that the $\Pphi/K$ ratios for \coll{A}{A}
collisions are only marginally enhanced with respect to their values
for \ppcoll\ collisions, and show relatively little variation as a
function of energy as seen in Fig.~\ref{fig:excit_phikp} and
Fig.~\ref{fig:excit_phikm}. Our parameterization of the \ppcoll\ cross
section for \Pphi\ production suggests that the $\Pphi/\PKm$ ratio
increases as \sNN\ approaches the threshold value from above for the
reaction $p p \rightarrow p p \Pphi$. The $\Pphi/\PKm$ ratios in
heavy-ion data seem to follow the energy dependence observed in
\ppcoll\ collisions rather reasonably.

\begin{figure}
\includegraphics[width=0.48\textwidth]{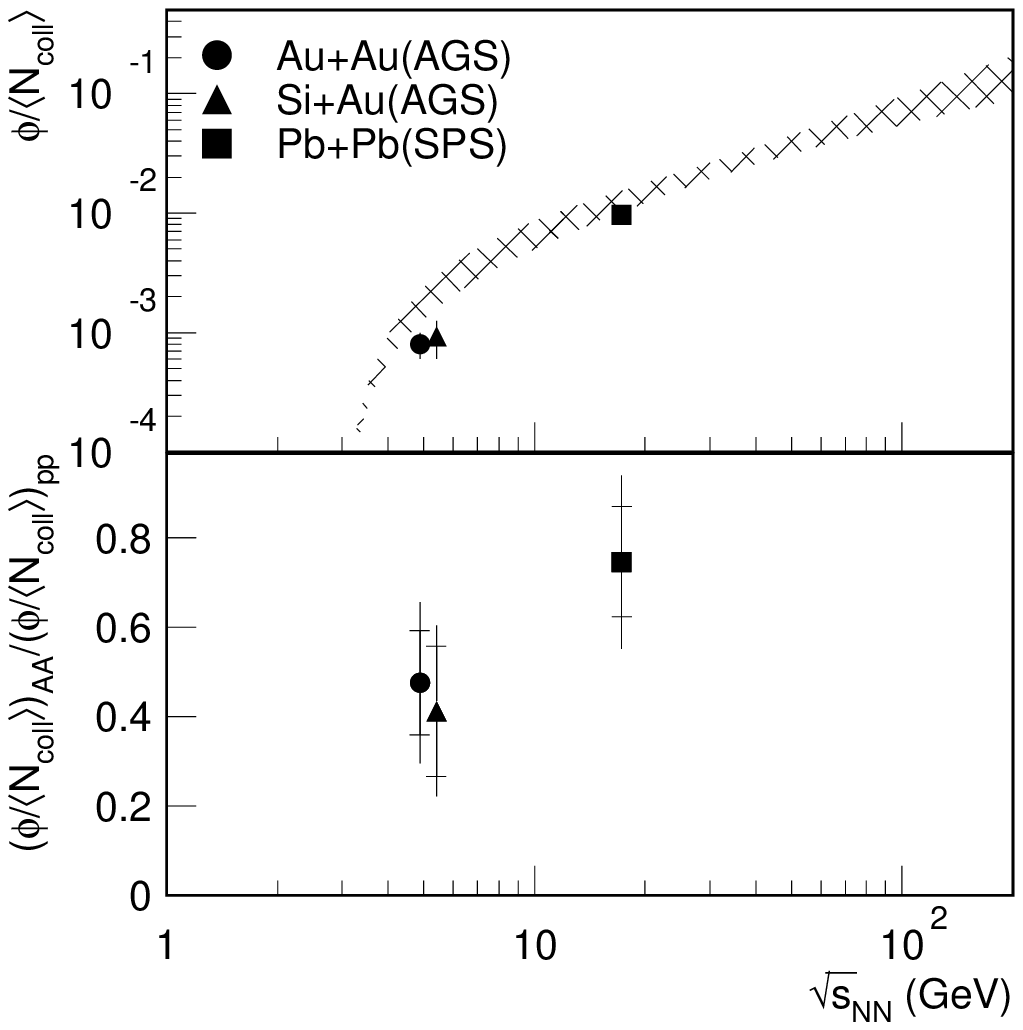}
\caption
{Same as Fig.~\ref{fig:excit_phipi} for the ratios of
$\Pphi/\mean{\ncoll}$, where \mean{\ncoll} is the number of binary
collisions. For the E917 data point, an estimated total \Pphi\ yield is
plotted as described in the text.}
\label{fig:excit_phinc}
\end{figure}

It is interesting to see if the data points from other systems and
energies obey the scaling with \mean{\ncoll} as well. The ratio
$\Pphi/\mean{\ncoll}$ for central \coll{A}{A} collisions at the AGS
and SPS is plotted in Fig.~\ref{fig:excit_phinc}. For the comparison,
we are forced to assume a value for the \Pphi\ rapidity width for our
measurement. A width of $\sigma_y = 0.71$ equal to the measured \PKm\
width is used, as explained in the previous section, and we plot our
point with an additional 20\% systematic uncertainty corresponding to
a range of values for $\sigma_y$ from 0.6 (overlap of \PKp\ and \PKm\
rapidity distributions) to 0.9 (\PKp\ rapidity distribution). The RHIC
point is not included in this comparison, because there is no
reasonable way to extrapolate the fiducial yield to all of phase
space. The ratio
$(\Pphi/\mean{\ncoll})_{\coll{A}{A}}/(\Pphi/\mean{\ncoll})_{\ppcoll}$
is consistent with \about{0.5}--0.7 for all three points from the AGS
and SPS. Our comparison is by no means precise. However, it does seem
that the yield of \Pphi's per binary collision, modulo the effects of
threshold and center-of-mass energy dependence, is approximately
constant across the three collisional systems studied. Furthermore,
instead of there being an enhancement in the \Pphi\ yield per binary
collision in heavy-ion collisions, this double ratio is less than 1.
This might reflect the effect of \Pphi\ absorption in the nuclear
fireball, as the inelastic cross section of the \Pphi\ on nucleons
(\about{8}--10~\mb) is a significant component of the total
interaction cross section
(\about{8}--12~\mb)~\cite{PHICROSS,PHICROSSABSORP}.

\section{Summary}
In conclusion, we have studied \Pphi\ production in \coll{Au}{Au}
collisions at 11.7\AGeVc\ around midrapidity as a function of
collision centrality. The yield per projectile participant shows a
steady rise toward central collisions. This enhanced production in
central collisions is stronger than that of non-strange $\pi$ mesons
as seen from the increasing $\phi/\pi$ ratio with centrality. The
ratios $\Pphi/\PKp$ and $\Pphi/\PKm$ are approximately constant with
\mean{\npp}. The yield of \Pphi's, like the yields of \PKp\ and \PKm,
is seen to scale with \mean{\ncoll}, the number of binary collisions,
and this observation is incompatible with predictions from 
some rescattering
models of \Pphi\ production, in which \Pphi\ production increases faster 
than kaon production with centrality due to combinatorial considerations.
Finally the yield of \Pphi's per binary
collision in \coll{A}{A} collisions is about 50--70\% of the \Pphi\
yield in \ppcoll\ collisions at AGS and SPS energies. That the yield
of \Pphi's per binary collision in \coll{A}{A} collisions is smaller
than that in \ppcoll\ collisions might signal the effect of \Pphi\
absorption by nucleons in heavy-ion collisions.

\acknowledgments
This work has been supported by the U.S. Department of Energy under
contracts with ANL (No.\ W-31-109-ENG-38), BNL (No.\
DE-AC02-98CH10886), MIT (No.\ DE-AC02-76ER03069), UC Riverside (No.\
DE-FG03-86ER40271), UIC (No.\ DE-FG02-94ER40865), and the University
of Maryland (No.\ DE-FG02-93ER40802); by the National Science
Foundation under contract with the University of Rochester (No.\
PHY-9722606); and by the Ministry of Education and KOSEF (No.\
951-0202-032-2) in Korea.

\appendix

\section{Parameterization of \Pphi\ production as a function of $\sqrt{s}$ in \ppcoll\ collisions}
\label{appendix}

The \Pphi\ yield in \ppcoll\ collisions provides a useful benchmark
for the interpretation of \Pphi\ yields in heavy-ion collisions.
There are various existing measurements of the inclusive total cross
section for \Pphi\ production in \ppcoll\ collisions for
$5<\sqrt{s}<60$~\GeV~\cite{NA49PHI,DISTO,PHIPP} that can be
extrapolated to the values of \sNN\ for which heavy-ion data exist.
(Note that the point at $\sqrt{s} = 2.90~\GeV$ from Ref.~\cite{DISTO},
while technically an exclusive measurement in the $p p \rightarrow
p p \PKp \PKm$ channel, is also an inclusive measurement, since no
other channels with a \Pphi\ or \PKm\ in the final state are
kinematically allowed at this energy.)  We have fit these
measurements with a form used to parameterize results on vector-meson
production cross sections in \ppcoll\ collisions obtained from
one-pion exchange calculations~\cite{SIBIRTSEV} and the Lund String
Model~\cite{CASSING}:

\begin{eqnarray} 
\sigma (pp \rightarrow \phi X) & = &  a(1 - x)^b x^c \nonumber\\
x & \equiv & s_\mathrm{thresh}/s  \nonumber\\
s_\mathrm{thresh} & = & 8.38~\GeV^2
\end{eqnarray}

\begin{figure}
\includegraphics[width=0.48\textwidth]{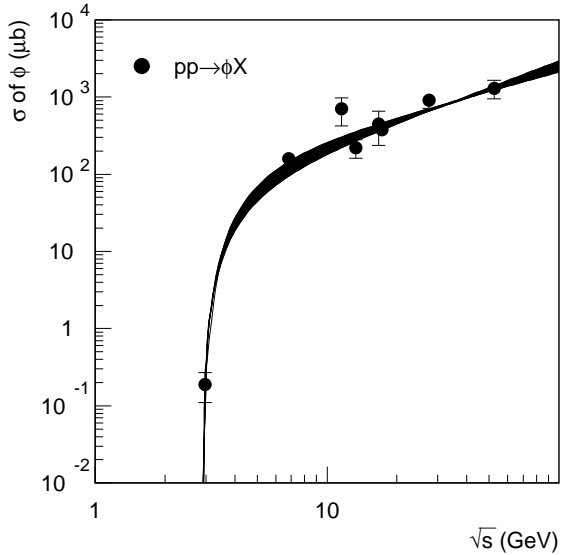}
\caption
{Parameterization of the \Pphi\ yield in $p+p$ interactions.
The data points are from Refs.~\cite{NA49PHI,PHIPP,DISTO}.
The fit is discussed in the text.}
\label{fig:pp_par}
\end{figure}

The results of our fit are shown in Fig.~\ref{fig:pp_par}.
We obtain $a = 74^{+36}_{-24}~\mu\mathrm{b}$, 
$b = 2.05^{+0.22}_{-0.18}$, and $c = 1.56^{+0.30}_{-0.27}$.
For comparison with heavy-ion data, we obtain yields by dividing the \Pphi\ 
cross sections for each value of \sNN\ by the inelastic \ppcoll\ 
cross section. For \coll{Au}{Au} collisions at the AGS, 
$\sNN = 4.87$~\GeV, and our parameterization of the \ppcoll\ data
gives a yield of $0.00169 \pm 0.00042$ at this energy.

For $K$ and $\pi$ production in \ppcoll\ reactions, we use the
multiplicity parameterizations in Ref.~\cite{KPIPP}. The smallest
$\sqrt{s}$ for the ranges of parameterization for $\pi$, \PKp and
\PKm\ are 3.0, 2.98, and 5.03~\GeV, respectively.

\end{document}